\documentclass[letter, longauth]{aa} 
\usepackage{txfonts}
\usepackage{graphicx}
\usepackage{rotating}
\usepackage{natbib}
\usepackage{lscape}
\usepackage{multirow}
\bibpunct{(}{)}{;}{a}{}{,} 
\newcommand{\ha}{H$\alpha$}

\newcommand{\brg}{Br$\gamma$}

\newcommand{\feii}{[\ion{Fe}{ii}]}

\newcommand{\oi}{[\ion{O}{i}]}

\newcommand{\av}{$A_V$}
\newcommand{\kms}{km\,s$^{-1}$}

\newcommand{\msun}{M$_{\odot}$}

\newcommand{\msunyr}{M$_{\odot}$\,yr$^{-1}$}

\newcommand{\macc}{$\dot{M}_{acc}$}
\newcommand{\mloss}{$\dot{M}_{jet}$}
\newcommand{\mratio}{$\dot{M}_\mathrm{jet}/\dot{M}_\mathrm{acc}$}
\bibpunct{(}{)}{;}{a}{}{,} 

\begin{document}

\title{Sub-0.1\arcsec \,optical imaging of the Z CMa jets with SPHERE/ZIMPOL}
\author{S. Antoniucci$^1$, L. Podio$^{2}$, B. Nisini$^{1}$, F. Bacciotti$^{2}$, E. Lagadec$^{3}$, E. Sissa$^{4,5}$, A. La Camera$^{6}$, T. Giannini$^{1}$, H. M. Schmid$^{7}$, R. Gratton$^{4}$, M. Turatto$^{4}$, S. Desidera$^{4}$, M. Bonnefoy$^{8,9}$, G. Chauvin$^{8,9}$, C. Dougados$^{8,9}$, A. Bazzon$^{7}$, C. Thalmann$^7$, M. Langlois$^{10,11}$}

\institute{$^1$ INAF-Osservatorio Astronomico di Roma, via Frascati 33, 00078 Monte Porzio Catone, Italy\label{oar}\\
$^2$ INAF-Osservatorio Astrofisico di Arcetri,  Largo E. Fermi 5, 50125 Firenze, Italy \label{oaa}\\
$^3$ Laboratoire Lagrange (UMR 7293), UNSA, CNRS, Observatoire de la C\^ote d'Azur, Bd de l'Observatoire, 06304 Nice Cedex 4, France\\
$^4$ INAF-Osservatorio Astronomico di Padova, Vicolo dell'Osservatorio 5, 35122 Padova, Italy\\
$^5$ Dipartimento di Fisica e Astronomia, Universit\`a degli Studi di Padova, Vicolo dell'Osservatorio 3, 35122 Padova, Italy,\\
$^6$ Dipartimento di Informatica, Bioingegneria, Robotica e Ingegneria dei Sistemi (DIBRIS), Universit\`a di Genova, via Dodecaneso 35, 16145, Genova, Italy\\
$^7$ Institute for Astronomy, ETH Z\"urich, Wolfgang-Pauli-Strasse 27, 8093 Z\"urich, Switzerland\\
$^8$ Universit\'e Grenoble Alpes, IPAG, F-38000 Grenoble, France\\
$^9$ CNRS, IPAG, F-38000, Grenoble, France\\
$^{10}$ Universit\'e Lyon, Universit\'e Lyon1, Ens de Lyon, CNRS, CRAL UMR5574, F-69230, Saint-Genis-Laval, France\\
$^{11}$ Aix Marseille Universit\'e, CNRS, LAM - Laboratoire d'Astrophysique de Marseille, UMR 7326, 13388, Marseille, France\\
}
\date{}

\abstract
{Crucial information on the mass accretion-ejection connection in young stars can be obtained from high spatial resolution images of jets in sources with known recurrent accretion outbursts.}  
{Using the VLT/SPHERE ZIMPOL instrument, we observed the young binary Z CMa that is composed of a Herbig Be star and a FUor object, both driving a jet. We aim to analyse the structure of the two jets, their relation with the properties of the driving sources, and their connection with previous accretion events observed in this target.
}
{We obtained optical images in the \ha\ and \oi\ 6300$\AA$ lines at the unprecedented angular resolution of $\sim$0.03 arcsec, on which we have performed both continuum subtraction and deconvolution, thereby deriving results that are consistent with each other.}
{Our images reveal extended emission from both sources: a fairly compact and poorly collimated emission SW of the Herbig component and an extended collimated and precessing jet from the FUor component. 
The compact emission from the Herbig star is compatible with a wide-angle wind and is possibly connected to the recent outburst events shown by this component. 
The FUor jet is traced down to 70 mas (80 AU) from the source and is highly collimated with a width of 26-48 AU at distances 100-200 AU, which is similar to the width of jets from T Tauri stars. This strongly suggests that the same magneto-centrifugal jet-launching mechanism also operates in FUors.
The observed jet wiggle can be modelled as originating from an orbital motion with a period of 4.2 yr around an unseen companion with mass between 0.48 and 1 \msun.
The jet mass loss rate \mloss\ was derived from the \oi\ luminosity and comprises of  between $1\times 10^{-8}$ and $1\times 10^{-6}$ \msunyr.
This is the first direct \mloss\ measurement from a jet in a FUor. 
If we assume previous mass accretion rate estimates obtained through modelling of the accretion disk, the derived range of \mloss\ would imply a very low mass-ejection efficiency (\mloss/\macc\ $\lesssim 0.02$), which is lower than that typical of T Tauri stars.}
{} 

\authorrunning{S. Antoniucci et al.}
\titlerunning{Sub-0.1\arcsec \,optical imaging of Z CMa}
\maketitle
%

\section{Introduction}
\label{sec:intro}

Star formation occurs through a mechanism that couples accretion from a disk and ejection of matter
in the form of winds and collimated jets, which concur in removing angular momentum from the system.
The accretion process is considered to be quasi-stationary, but time-variable accretion bursts caused 
by a quick rise of the mass accretion rate from the disk are sometimes observed.
Studies of jets from sources that show episodic accretion events allow one to
trace the past history of the accretion bursts, as one may expect to detect spatial structures
within the jet that can be connected with particularly intense mass-ejection events.
Fundamental insights come from the inner regions close to the source, where
the jets are less contaminated by interactions with the ambient medium, so that
their morphology can be tightly connected with variations of the ejection efficiency.

The Z CMa binary is a peculiar system that is suited to investigate the occurrence of 
mass ejection in young sources undergoing episodic accretion. 
This binary is composed of a Herbig Be star (the NW component), which shows EXor-like outbursts \citep{szeifert10},
and a second component (SE) that exhibits broad double-peaked optical absorption lines that are typical of a FU Ori object 
\citep{hartmann89,hartmann96}.
The two sources, located at a distance of 1150 pc \citep{herbst78}, are separated by 0\,\farcs1 and only recently high angular resolution 
observations with adaptive optics (AO) systems and interferometers have 
allowed the separatele determination of the properties of the two sources
(Bonnefoy et al. 2016, submitted; Hinkley et al. 2013). \nocite{hinkley13}
The Herbig star, which is responsible for the variability of the binary (Bonnefoy et al. 2016, submitted), 
has shown recurrent variations in the visual magnitude of 2-3 mag in the last ten years with recent major events 
recorded in 2008, 2011, and 2015 \citep[see light curves in][]{canovas12, maehara15}.
These variations have been imputed to both variable accretion and to changes of the extinction along the line of sight \citep{benisty10, szeifert10}.

An extended optical jet from Z CMa was originally detected by \citet{poetzel89}.
Recently, \citet[][hereafter W10]{whelan10} discovered that both sources drive 
their own jet, at slightly different position angles (P.A.): 245$\degr$ (east of north) for the Herbig and 235$\degr$ for the FU 
Ori. The jet from the NW source has a very high radial velocity reaching up to $-$600\kms\, 
while the jet from the SE component has a velocity of $\sim$ $-$200/$-$300 \kms. 
The large-scale optical jet observed by \citet{poetzel89} is usually associated 
with the Herbig component \citep[e.g.][W10]{garcia99}. 
Additional emission that is not collimated well   is observed at intermediate velocities, which  cannot however be 
unambiguously associated with one of the two components (W10).
The analysis of these jets offers us the opportunity to obtain 
direct information on the mass loss and indirect information on the accretion rate.
This is especially interesting in the case of the FUor component, since no direct mass loss measurements from jets are available for such objects.
Wind mass loss rates as high as $10^{-5}$ \msunyr were estimated on the prototype source FU Ori from line profiles \citep{croswell87,calvet93}. 
Accretion rates estimates for FUor objects are also very scarce and range from  $10^{-4}$ to $10^{-6}$ \msunyr \citep{audard14}.

We present here VLT/SPHERE \citep{beuzit08} high angular resolution optical observations of 
Z CMa in the \oi6300\AA\, and H$\alpha$ lines. The extreme AO performance of SPHERE
allows us to have direct images of the Z CMa system at unprecedented high 
contrast and spatial resolution ($\sim$0\,\farcs03).

\vspace{-0.3cm}
\section{Observations and data reduction}
\label{sec:obs_red}

Observations of Z CMa in the \ha\ and \oi\ lines were performed using the ZIMPOL instrument 
\citep{thalmann08,schmid12} of SPHERE on the night of 31 March 2015.
Simultaneous images of Z CMa in \ha\ and in the adjacent continuum were acquired with the two ZIMPOL cameras using 
the B\_Ha ($\lambda_{c}$=655.6 nm, $\Delta\lambda$=5.5 nm) and Cnt\_Ha ($\lambda_{c}$=644.9 nm, $\Delta\lambda$=4.1 nm) filters. 
Exposures were taken in field-stabilized mode using two different field rotation angles and swapping the filters in front of each camera to optimize artefact removal during the reduction. This resulted in 60 frames per filter (15 for each setup) for a total exposure time of 30 minutes.
Images in the \oi\ line with the OI\_630 filter\footnote{Simultaneous observations on the continuum are not possible with this filter, which is mounted on the common wheel of the two cameras.} ($\lambda_{c}$=629.5 nm, $\Delta\lambda$= 5.4 nm) were obtained in field-stabilized mode using two different field rotation angles for a total of 30 frames and an integration time of 30 minutes.
Atmospheric conditions remained stable during observations (seeing $\sim$ 0\,\farcs8-1\arcsec, $\tau_0 \sim 2.5$ ms).

The raw data were processed using the SPHERE-ZIMPOL IDL pipeline (version 1.1) developed at ETH Z\"urich, 
which performs bias, dark, flat-field correction, re-centring of the dithered frames, and allows for derotation, selection, subtraction, and average or median combination of the frames.
The Cnt\_Ha exposures were used to remove the continuum emission from the B\_Ha and OI\_630 filter images, so as to 
produce continuum-subtracted \ha\ and \oi\ images in which the line-emitting structures around the components are evidenced.
Details about the reduction procedure for the continuum subtraction are given in Appendix A.
From the FWHM of the PSF of the two stars measured in the Cnt\_Ha frames we estimate an effective angular resolution of $\sim$30 mas for our images. The binary separation and position angle are analysed in Appendix B.

We also performed an alternative processing of the ZIMPOL images with the multi-component Richardson-Lucy (MC-RL) deconvolution method developed by \citet{la_camera14}, which is optimized for the reconstruction of high dynamical range images, such as those of jets from young stars \citep[e.g.][]{antoniucci14c}. The MC-RL method is able to separately reconstruct the star(s) and the underlying diffuse emission in the image.
We extracted the PSF from the Cnt\_Ha image via the blind deconvolution algorithm described in \citet{prato13,prato15} and employed to reconstruct both the \oi\ and \ha\ images.

\vspace{-0.3cm}
\section{Results and discussion}
\label{sec:results}

The final continuum-subtracted \oi\ and \ha\ images are shown in the upper part of Fig.~\ref{fig:images}a,b. 
We can recognise two major structures: an extended collimated jet from the FUor component and a more
compact and poorly collimated emission SW of the Herbig star. 
These structures are especially evident in the \oi\ image, which is much less affected by subtraction residuals than the \ha\ image. 
In our \oi\ continuum-subtracted image we can trace the flows down to about $\sim$90 mas from the sources 
(corresponding to $\sim$105 AU). 
The morphology of both the FUor jet and the compact emission from the Herbig is thoroughly 
confirmed by the deconvolved images (lower part of Fig.~\ref{fig:images}a,b), which 
represent the diffuse emission in the image as reconstructed by the MC-RL algorithm.
The \oi\ deconvolved image shows details of the FUor jet even closer to driving source, down to $\sim$70 mas (80 AU).
The deconvolved \ha\ image provides evidence of the non-collimated emission SW of the Herbig star, which was not fully visible 
in the continuum-subtracted image owing to the large residuals.

\vspace{-0.3cm}
\subsection{Outflow from the Herbig Be star}
\label{sec:jet_H}

The extended \oi\ and \ha\ emission detected SW of the Herbig star is fairly compact 
and poorly collimated, suggesting an origin in a wide-angle wind more than in a jet. 
The flow has an extent of $\sim$0\farcs12, corresponding to $\sim$140 AU, while the 
observed opening angle is $\sim55\degr$ with a median P.A. around 230$\degr$. 
In \feii\ KECK observations by W10, collimated emission travelling at velocities
up to $-$600 \kms and with a P.A.$\sim$245$\degr$ is detected
up to distances of 0\farcs4 from the source, along with a more diffuse emission at lower velocities (100-200 \kms).
The wide-angle compact emission that we observe is likely connected to this diffuse emission, while
the high-velocity collimated jet is not visible in our images. 
Although the Herbig jet is visible at high velocities in the \feii\ data reported by W10 (which are not flux calibrated), the 
flux contrast between the FUor and Herbig jet is indeed greater than a factor $\sim$20 for velocities between $-$400 and $-$90 \kms. This also agrees
with the non-detection of the jet in the SINFONI data shown by W10.
As we can estimate a mean signal-to-noise ratio of about 12 on the FUor jet, we conclude that we did not observe the collimated 
jet from the Herbig because of sensitivity limits.

Interferometric measurements of the \brg\ line during the 2008 outburst \citep{benisty10} suggested a bipolar (non fully spherical) wind 
from the star on mas scales, roughly aligned in the direction of the large-scale jet. As the \brg\ emission disappeared outside the outburst 
phase, the authors concluded that the wind emission is connected to the event of increased mass accretion driving the outburst.
To understand if the emission detected in our images is related to such bipolar wind,
we evaluated whether the signal may be attributed to an ejection event from the recent Herbig star bursts, 
which occurred in January 2008 and February 2011 \citep{canovas12}, and more recently in January 2015 \citep[see photometry from
the Kamogata Wide-field Survey][]{maehara15}. 
Given the observed extent of the flow, we find that it should travel at a tangential velocity of 90, 170, and
2700 \kms to have been ejected from the 2008, 2011, or 2015 events, respectively.
Ruling out the latter event, which implies an exceedingly large velocity, a connection with the 2008 and 2011 events is possible.

\vspace{-0.3cm}
\subsection{Jet from the FUor}
\label{sec:jet_F}

\begin{figure*}[t]
\centering
\includegraphics[width=4.4cm]{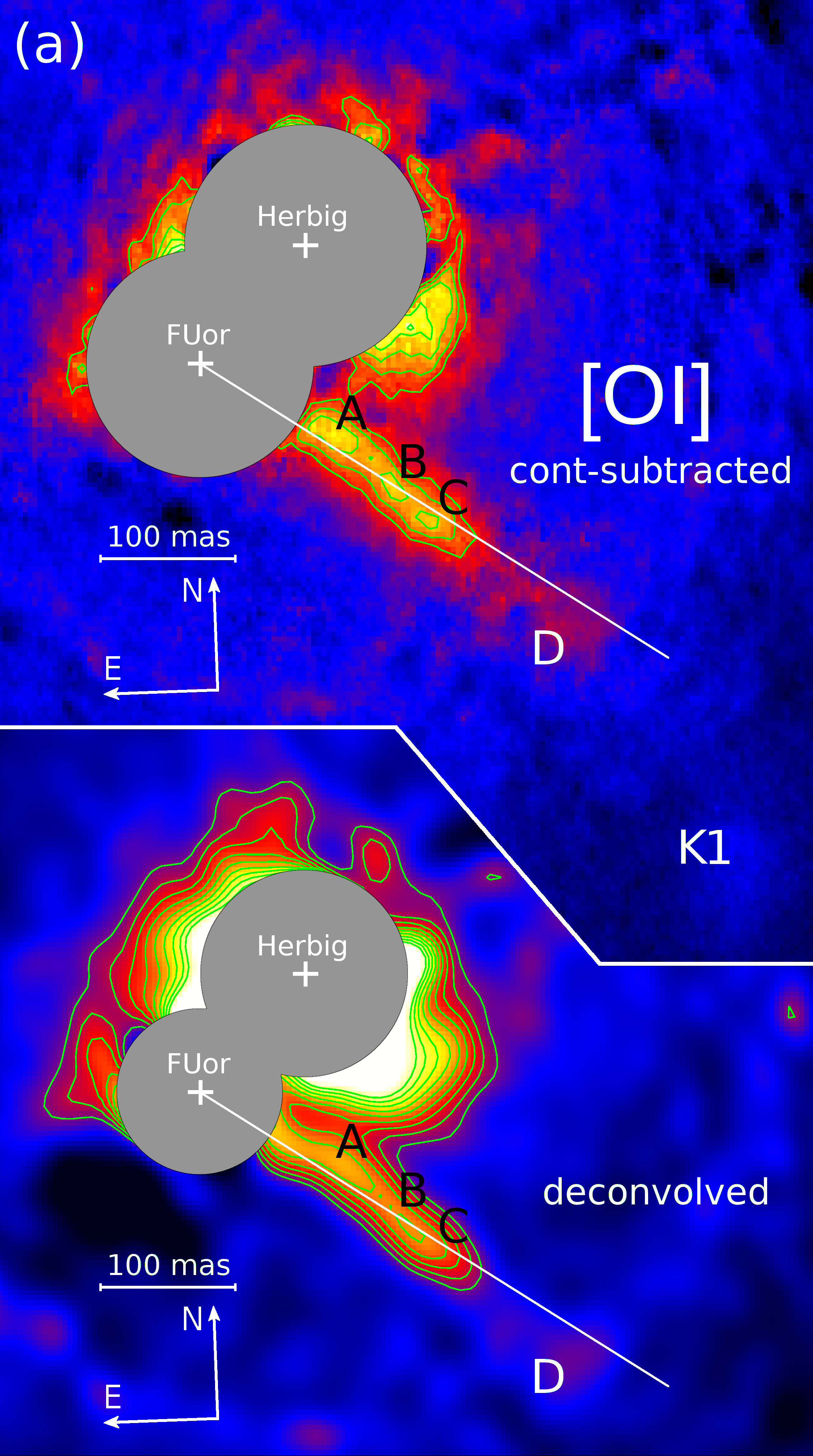}
\includegraphics[width=4.4cm]{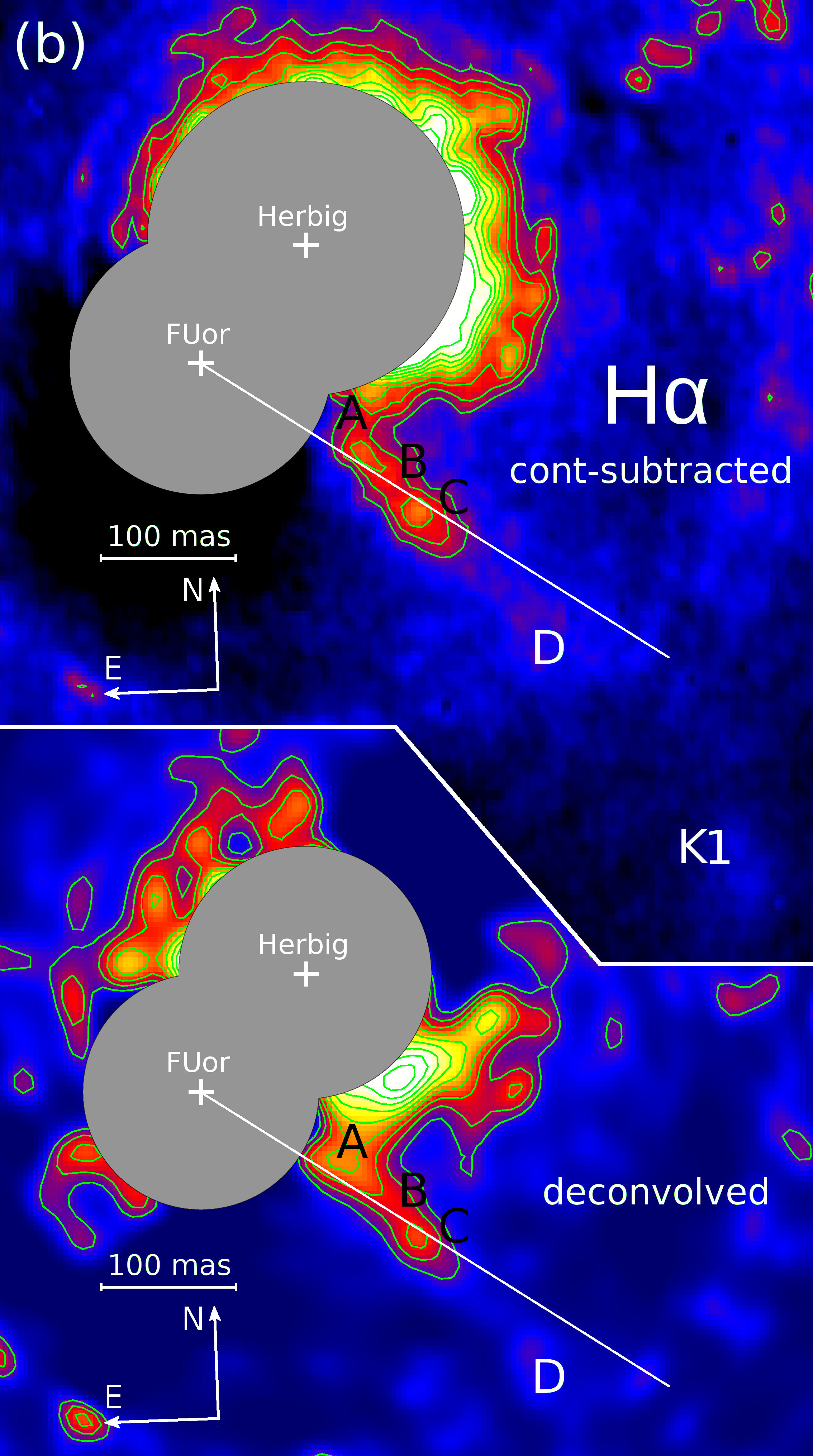}
\includegraphics[width=5.7cm]{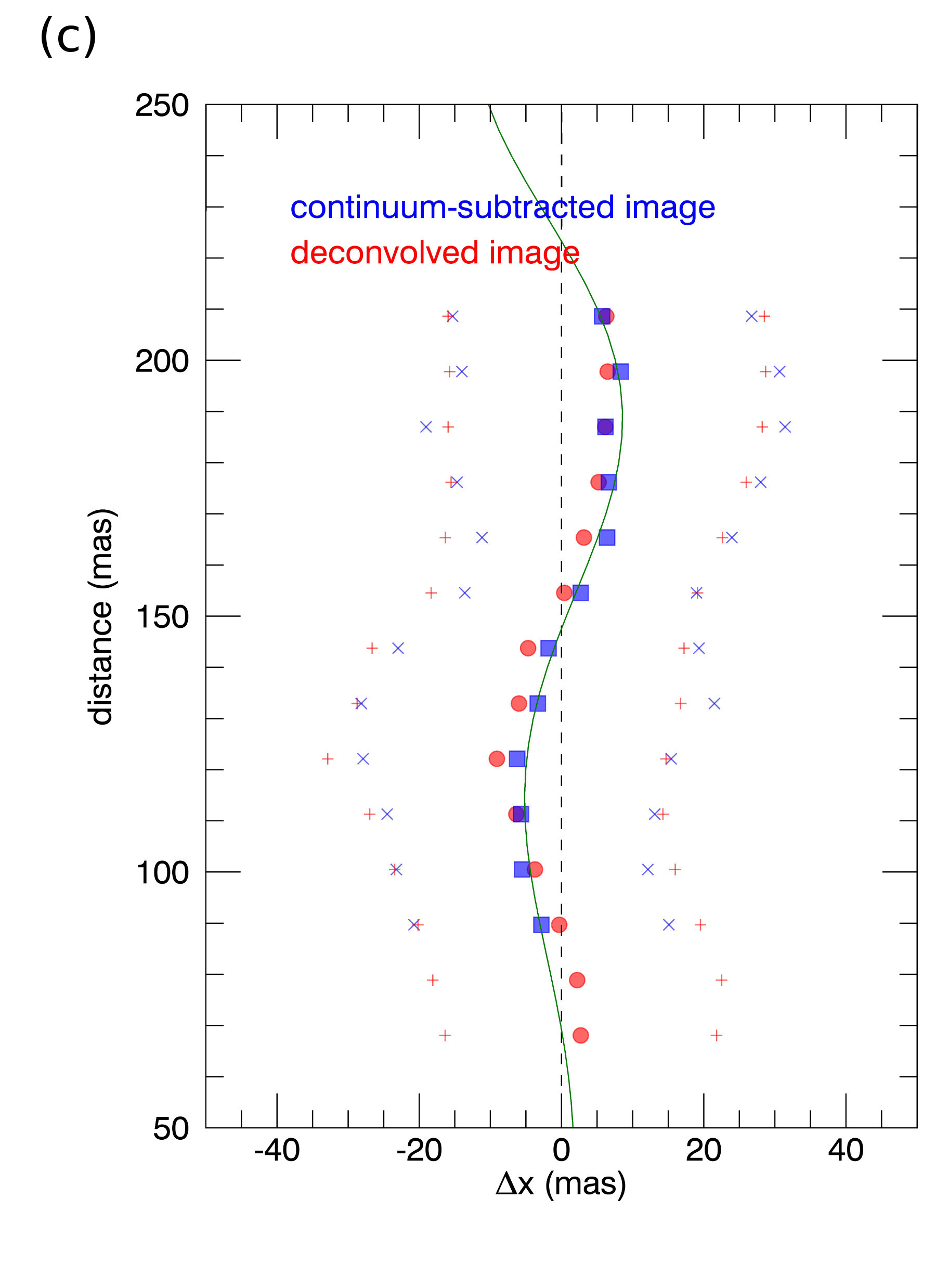}

\caption{\label{fig:images} 
\textbf{(a)}: Continuum-subtracted (top) and deconvolved (bottom) \oi\ images of the Z CMa system. The deconvolved image shows only the diffuse emission as reconstructed by the MC-RL algorithm. The centroids of the two stars and the position of the main knots of the FUor jet are indicated. Areas heavily corrupted by artefacts around the star centroids have been masked. 
\textbf{(b)}: Same for the \ha\ observations. 
\textbf{(c)}: Fitted transverse positions of the peak of the FUor jet spatial profile as a function of the distance from the exciting source for both the \oi\ continuum-subtracted (blue squares) and deconvolved images (red circles). Blue crosses and red pluses indicate the measured profile width of the two images. The solid green line is the best fit of the wiggle produced by an orbital motion of the jet source around a companion \citep{anglada07} to the peaks of the continuum-subtracted image. 
}
\end{figure*}

The collimated jet from the FUor object can be identified with the Jet B observed in \feii\ by W10, which travels
at radial velocities between $-$100 and $-$400 \kms. 
The morphology of the jet in our \oi\ image is however much more structured than in the \feii\ images.
In the continuum-subtracted image, we can identify three peaks on the brightest part of the jet, which we name A, B, and C, 
while we denote with D a more diffuse nebulosity located farther out (Fig.\ref{fig:images}a,b).
The section of the jet between knots A and C shows a very apparent wiggling, which we analyse in detail in Sect.~\ref{sec:wiggling}.
Outside the ring of enhanced speckle noise (with radius $\sim$0\farcs3),
we also detect a weak emission knot that we identify as the knot K1 
of W10. As in the \feii\ images, this knot is not aligned 
with the rest of the jet although it is probably connected with it. The knot 
shows an appreciable proper motion with respect to the W10 images. 
In the assumption that \feii\ and \oi\ trace the same gas, we measure an increase of 186 mas for the knot distance from the source in the two images,
which were taken 5.3 years apart. This corresponds to a tangential velocity of 192 km\,s$^{-1}$. 
Hereafter we assume that the rest of the jet has the same tangential velocity. 
With this assumption, we can infer the times at which the A-D knots were ejected, obtaining  
dynamical timescales of 3.8, 5.4, 6.0, and 9.3 yr for knots
A, B, C, and D, respectively. We therefore expect that
only knot D should be visible in the W10 \feii\ image. 
Indeed, only one knot is clearly identifiable in that image and its dynamical timescale
is consistent with it being knot D of our image.
The derived tangential velocity combined with the observed radial velocities between $-$100 and $-$400 \kms\ implies 
a jet inclination angle with respect to the plane of the sky in a range between 28\degr\ and 64\degr.

\vspace{-0.3cm}
\subsubsection{Jet wiggling}
\label{sec:wiggling}

From the observed wiggling we can estimate a mean P.A. of the jet of 236.5$\degr$, which approximately intersects the knot B, and 
a variation of about $\pm$3$\degr$ around such direction.
We analysed the jet wiggling in our \oi\ image by considering 12 contiguous 3-pixel-wide slices orthogonal to the jet axis 
covering the section between knots A and C, where we have the best signal to noise. 
We averaged the counts within each slice to obtain the spatial profiles of the jet at various distances. 
Using a linear combination of a second degree polynomial (to fit the local continuum) and a Gaussian function, we 
computed the profile peak positions as a function of the distance from the star, which are reported in 
Fig.~\ref{fig:images}c for both the subtracted and deconvolved \oi\ image.

The jet profiles are barely resolved with fitted full width and half maxima (FWHM) of the Gaussians in the continuum-subtracted image 
spanning between 35 and 50 mas with a general trend of increasing width with distance (Fig.~\ref{fig:images}c). 
When corrected for the instrumental profile, these indicate a jet intrinsic FWHM in the range 26-48 AU at distances 
100-200 AU from the star; these values
are in substantial agreement with widths observed in Class II jets and also in less 
evolved sources \citep{cabrit07} and testifies a high degree of collimation of the jet.  
This finding is an important indication that the same magneto-centrifugal jet-launching mechanism is probably at 
work in objects with high accretion rates such as FUors with their massive peculiar disks.

With a jet tangential velocity of 192 \kms, the observed wiggling occurs on timescales of a few years, 
which suggests an origin due to the interaction with a close undetected companion.
We therefore explored the hypothesis that the wiggling arises from the orbital motion of the jet-emitting source around this companion.
We adopt the formulation of \citet{anglada07}, which considers a ballistic jet from a star in a circular orbit with a constant ejection 
velocity orthogonal to the orbital plane. Using Eq.~8 of \citet{anglada07}, which gives the observed jet shape on sky as a function of the 
jet half-opening angle $\alpha$ and of the projected angular period of the wiggles $\lambda$, we fit the peak positions measured in the continuum-subtracted image (Fig.~\ref{fig:images}c) and obtain  
the best-fit parameters $\alpha=2.6\degr \pm 0.2\degr$ and $\lambda=148.8 \pm 8.3$ mas. These values indicate an absolute orbital radius of 
the jet source $r_o=1.2 \pm 0.2$ AU and, for a jet tangential velocity of 192 \kms, an orbital velocity $v_o = 9 \pm 1 $ \kms and a 
orbital period $t_o = 4.2 \pm 0.8$ yr. 
The fitted wiggling is also roughly compatible with the position of the more external and diffuse knot D, whereas the location of the farther K1 knot deviates by more than 50 mas from the expected position. However, at larger distances from the sources we expect that the interaction of the jet with the ambient medium may produce significant deviations of the jet stream.
In the adopted formulation the total mass of the binary system is given by $M/$\msun$ = \mu^{-3} (r_o/$AU$)^3 (t_o/$yr$)^{-2}$ $\simeq 0.107 \mu^{-3}$, where $\mu = m_2/M$ is the ratio of the mass of the companion to the total mass. 
Previous estimates of the FUor mass range between 1 and 3 \msun\, (Hartmann et al. 1989, 1986, Van den Ancker 2004). If we consider this as the possible range for the total mass of the binary, we obtain $m_1$(mass of the jet source)=0.52 \msun\, $m_2$=0.48 \msun\ and $m_1$=2.0 \msun\, $m_2$=1.0 \msun\ for the lower and upper limit, respectively. The expected separation would be between 3.8 and 2.6 AU, corresponding to 3.3-2.3 mas, which is well below our angular resolution.
This scenario is compatible with the binary hypothesis considered by \citet{millan-gabet06} to explain the low interferometric visibility measured in the $K$ band. 
No signature of such a binary was visible in the optical spectra of \citet{hartmann89}, which had a resolution of 12 \kms and were however dominated by disk emission.

\vspace{-0.3cm}
\subsubsection{Jet mass ejection rates}
\label{sec:mloss}

Further clues to the launch of the FUor jet and the accretion of its driving source can be gathered by measuring the jet mass loss rate (\mloss).
In principle, both the \oi\ and \ha\ flux in the jet could be used for this purpose; however, the \ha\ emissivity is subject 
to a much higher uncertainty owing to the contribution of both recombination and collisional excitation \citep[see][]{bacciotti99}.
In addition, \ha\ could be optically thick at the jet base, therefore not tracing all
the jet mass. Therefore, we chose to consider only the \oi\ line for determining the \mloss.
To derive the \oi\ flux we first flux-calibrated the images following the procedure described in Appendix C.
The jet \oi\ flux was then measured in a rectangular box of 160~mas in length, which covers the part of the 
jet between knots A-C, and 85~mas in width. Four equivalent boxes on adjacent areas were used to evaluate the background contribution. 
Based on the standard deviation of the background counts in these areas we estimated a 10\% uncertainty on the jet count measurement.
The jet flux eventually derived is $1.1 \pm 0.3 \times10^{-14} $ erg s$^{-1}$ cm$^{-2}$. 
Estimates of the visual extinction towards the target span between 2.4 and 4.6 mag \citep[][and references therein]{stelzer09}, which suggest a relatively low circumstellar \av\ on source. Hence, we considered a  
purely interstellar extinction towards the jet and assumed a value \av = $1.0 \pm 0.5$ mag based on the $\sim$1 kpc distance of Z CMa \citep{savage79}. 
The final extinction-corrected flux is $2.3 \pm 1.5 \times10^{-14} $ erg s$^{-1}$ cm$^{-2}$.

From the line luminosity, we can compute \mloss\ by considering the projected jet length, 
the measured tangential velocity of 192 \kms, and the oxygen emissivities \citep[see e.g.][]{antoniucci08}, 
which depend on the gas temperature $T$ and electron density $n_e$.
For the computation, we considered an oxygen abundance of 4.6 $\times10^{-4}$ \citep{asplund05} and assumed that all oxygen is in neutral state.
The emissivities were taken from the grids of the the NEBULIO database\footnote{http://www.oa-roma.inaf.it/irgroup/line\_grids/Atomic\_line\_grids/\\Home.html} \citep[see also][]{giannini15}.
By adopting $n_e = 10^4$ cm$^{-3}$ and a gas temperature T=10\,000 K 
\citep[as typically measured in T Tauri jets; see][]{bacciotti99,maurri14}
we obtain \mloss\ $2.7 \pm 1.5 \times 10^{-7}$ \msunyr.
Allowing for a range of values between $5\times 10^{3}$ and 
$5\times 10^{4}$ cm$^{-3}$ for the electron density and 
between 8\,000 K and 12\,000 K for the temperature, we eventually obtain a possible range of \mloss\ 
between $1 \times 10^{-8}$ and $1 \times 10^{-6}$ \msunyr.
To our knowledge this is the first direct measurement of the mass loss 
rate from jet emission observed close to a driving FUor object.
A mass accretion rate \macc = $7.9\times 10^{-5}$ for the FUor component was estimated by \citet{hartmann96} based on the modelling of the accretion disk around the source and comparison with observations. If we assume this \macc\ value, the derived mass loss 
rate range would imply a very low ejection efficiency \mratio\ $\lesssim$0.02, which is lower than the typical values around 10\% found in T Tauri stars \citep{podio11,ellerbroek13}.
This is somewhat unexpected, as the jet collimation strongly suggests a common magneto-centrifugal ejection mechanism. Moreover, at the high 
accretion rates of the FUor, the surface of the disk is expected to be hotter than in classical T Tauri stars, which should 
favour the mass loading of the jet streamlines that arise from the disk, thus increasing the ejection efficiency. 
This suggests that the mass accretion rate provided by \citet{hartmann96} may be over-estimated.
Alternatively, the mass ejection mechanism in the FUor might be peculiar compared to T Tauri stars, despite the common high jet collimation, for instance because of disk instabilities possibly induced by the close companion.
Another explanation is that the collimated jet is tracing only a minor part of the ejected matter, while the bulk of the mass loss is due to wider angle winds \citep[e.g.][]{machida14}, which however remain undetected in our image.


\vspace{-0.3cm}
\bibliographystyle{aa} 
\bibliography{refs} 


\section*{Appendix A: Continuum-subtraction procedure}

For the \ha\ observations, we performed a frame-by-frame subtraction of the B\_Ha and Cnt\_Ha images to remove the continuum 
emission and evidence the line-emitting structures around the components.
Prior the subtraction, each Cnt\_Ha frame must be normalized to the corresponding B\_Ha frame to account for the different filter band-passes. The 
normalization factor is empirically computed from the ratio of the total counts measured in the same region of the two images. 
The comparison between the photon count ratios between the two Z CMa components in the continuum and in the \ha\ frames clearly reveals 
a strong \ha\ emission from the NW component, in agreement with previous observations of the separate spectra of the two stars \citep{hinkley13}. 
This makes it difficult to find a normalization that efficiently removes the intense continuum from both stars at the same time.
Indeed, if we normalize to the signal of the SE component, the point-like \ha\ emission from the Herbig star creates a very strong residual PSF signal 
that hampers the detection of the structures around both objects. Conversely, by normalizing to the Herbig component, we oversubtract the signal 
from the FUor and eventually create a deep residual negative region around the binary where no information can be retrieved. 
The best compromise is found with a normalization factor calculated over a region including both components. 
After de-rotating the observations taken at a different field position angle, we finally performed a median of all the subtracted 
frames.
 
For the \oi\ data, we produced a median \oi\ image from all the frames, after de-rotating those obtained at a different field position angle. Analogously, we created a median image of the Cnt\_Ha frames, which we normalized to the signal of the Herbig star and then subtracted from the median \oi\ image. The residuals from continuum subtraction are more compact for \oi, probably because the point-like \oi\ emission from the NW source is much weaker than \ha.

\section*{Appendix B: Binary separation and orbital motion}
\label{sec:orbit}

\begin{figure}[t]
\centering
\includegraphics[width=8.5cm]{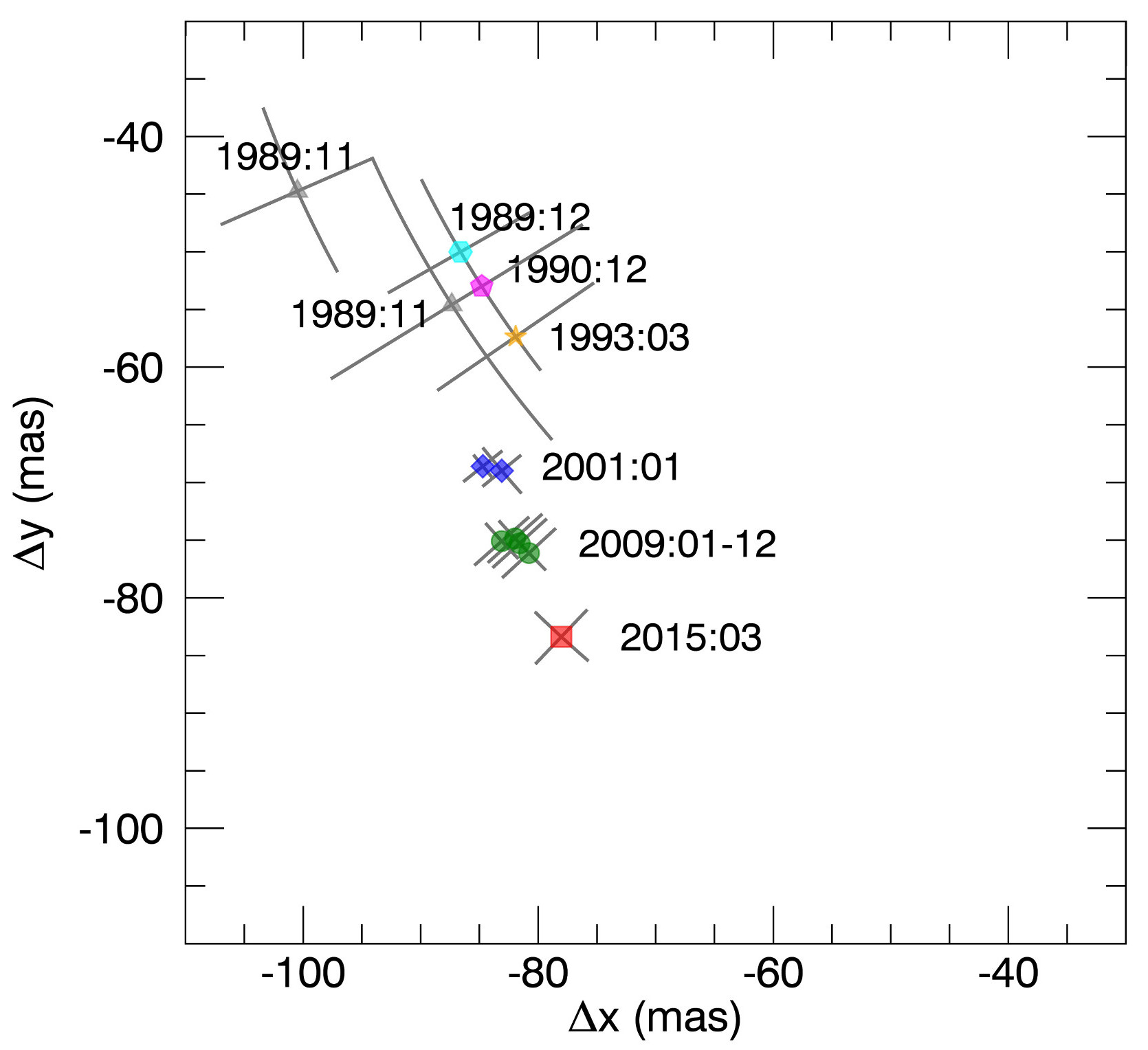}
\caption{\label{fig:orbit} 
Relative position of the FUor component with respect to the Herbig from measurements. Labels report year:month of observation (see text for the references). The red point refers to our SPHERE observations.
}
\end{figure}

We derived the separation and P.A. of the Z CMa components by computing the average centroids of the two stars from the 60 frames 
acquired with the Cnt\_Ha filter. 
Considering the ZIMPOL pixel scale of 3.6006 $\pm$ 0.0051 mas/pix and the vertical axis position angle of 358.39 $\pm$ 0.11 deg (Ginski et al., in prep.),
we obtain a separation of 114.2$\pm$3.1 mas and a position angle of 136.9$\pm$1.5 deg, which corresponds to a projected length of about 131 AU. This measurement and others collected in a time span of 26 years (Koresko et al. 1991; Haas et al. 1993; Thiebaut et al. 1995; Barth et al. 1994; Millan-Gabet et al. 2002; Bonnefoy et al. 2016, submitted) are reported in Fig.\ref{fig:orbit}. The plot provides evidence of the variation of the secondary component position relative to the Herbig star. Our measurement 
indicates that the separation is still increasing with the position angle, 
suggesting a (projected) elliptical shape. 
If we consider masses of 16 and 3\msun\ for the Herbig and FUor \citep{van_den_ancker04} and assume an inclined circular orbit, 
the still increasing separation indicates an orbital period longer than 345 yr. 
A more detailed analysis of the orbit is presented in \citep{bonnefoy16}.

\section*{Appendix C: Flux calibration of the \oi\ image}

To perform the flux calibration of the \oi\ image, we first calibrated the Cnt\_Ha image 
using the conversion coefficient reported by Schmid et al. 
(2016, in prep.), assuming a 10\% uncertainty on the provided factor. 
To check the goodness of this calibration, we derived an estimate of the binary integrated $R$-band magnitude by 
measuring the counts in a circular aperture including both components.
We thus obtain $R= 8.5 \pm 0.2$ mag for the binary. 
The measured magnitude is consistent with the integrated magnitudes of 
$V=9.3$ and $I_C=8.0$ reported by the Kamogata Wide-field Survey on 30 Mar 2015 \citep{maehara15}, i.e. 
the night before our SPHERE observations, also considering that the typical photometric variability of 
Z CMa on a timescale of one day is below 0.1 mag.

To flux calibrate the \oi\ image, we then made the assumption that the continuum flux density in the Cnt\_Ha
image (at $\lambda=645$ nm) is the same for the \oi\ image (at $\lambda=630$ nm). On this basis, 
we were able to derive a count-to-flux conversion factor for 
the \oi\ image by considering a circular area centred on the SE 
component on both images. 
We selected the FUor because we expect a weaker on-source \oi\ 
line emission on this component than on the Herbig star.
Using different radii for the circular aperture we derive conversion factors that differ for less than 2\%.
To be conservative, also considering the previous continuum flux density assumption, 
we associate a relative uncertainty of 10\% on the computed factor. 
This, in conjunction with the initial uncertainty on the Cnt\_Ha image conversion coefficient, allows us to evaluate an 
accuracy of 20\% for the flux calibration of the \oi\ image. 

\end{document}